\documentclass[fleqn,usenatbib]{mnras}

\usepackage{newtxtext,newtxmath}

\usepackage[T1]{fontenc}
\usepackage{ae,aecompl}

\usepackage{graphicx}	
\usepackage{amsmath}	
\usepackage{amssymb}	
\usepackage{calc}
\usepackage{ifthen}
\usepackage{array}
\usepackage{algorithmicx}
\usepackage{algpseudocode}

\newcounter{MCMCDone}
\def\MCMC{\ifthenelse{\equal{\arabic{MCMCDone}}{0}}{Markov Chain Monte Carlo (MCMC)\setcounter{MCMCDone}{1}}{MCMC}}

\title{A random-walk model for dark matter halo spins}
\author[A. J. Benson, C. Behrens, \& Y. Lu]{Andrew Benson\,$^{1}$\thanks{E-mail: abenson@carnegiescience.edu}, Christoph Behrens\,$^{2}$, Yu Lu\,$^{1}$\\
$^{1}$ Carnegie Observatories, 813 Santa Barbara Street, Pasadena, CA 91101, USA\\
$^{2}$ Institut f\"{u}r Astrophysik, Georg-August Universit\"{a}t G\"{o}ttingen, Friedrich-Hundt-Platz 1, 37077, G\"{o}ttingen, Germany\\
}

\begin{document}

\maketitle

\begin{abstract}
We extend the random-walk model of Vitvitska et al. for predicting the spins of dark matter halos from their merger histories. Using updated merger rates, orbital parameter distributions, and N-body constraints we show that this model can accurately reproduce the distribution of spin parameters measured in N-body simulations when we include a weak correlation between the spins of halos and the angular momenta of infalling subhalos. We further show that this model is in approximate agreement with the correlation of the spin magnitude over time as determined from N-body simulations, while it slightly underpredicts the correlation in the direction of the spin vector measured from the same simulations. This model is useful for predicting spins from merger histories derived from non-N-body sources, thereby circumventing the need for very high resolution simulations to permit accurate measurements of spins. It may be particularly relevant to modeling systems which accumulate angular momentum from halos over time (such as galactic discs)---we show that this model makes small but significant changes in the distribution of galactic disc sizes computed using the {\sc Galacticus} semi-analytic galaxy formation model.
\end{abstract}


\section{Introduction}

The angular momenta of dark matter halos has long been understood to arise through tidal torques acting on the proto-halo \citep{hoyle_origin_1949,peebles_origin_1969,doroshkevich_spatial_1970,white_angular_1984,barnes_angular_1987,porciani_testing_2002}. These torques also impart angular momentum to the baryonic component of forming halos, and this is later incorporated into galaxies. Simple models, based on the assumption that material collapsing to form galactic discs conserves its original angular momentum, predict sizes of galactic discs in approximate agreement with observations \citeauthor{fall_formation_1980}~(\citeyear{fall_formation_1980}; see also \citealt{mo_formation_1998}, but see \citealt{jiang_is_2018} who show that galactic angular momentum is in fact not well correlated with halo spin, at least at zero time lag).

The angular momentum of halos, typically characterized by the dimensionless spin parameter, $\lambda$, has been measured directly from N-body simulations \citep{cole_structure_1996,bett_spin_2007,gottlober_shape_2007,maccio_concentration_2007,zhang_spin_2009,lee_properties_2016,rodriguez-puebla_halo_2016,zjupa_angular_2017}. However, \citeauthor{benson_constraining_2017}~(\citeyear{benson_constraining_2017}; see also \citealt{trenti_how_2010}) showed that spins are often poorly determined in N-body simulations because of particle noise---a 10\% precision measurement of $\lambda$ requires at least 50,000 bound particles in a halo. As the majority of halos found in cosmological N-body simulations will contain far fewer particles than this, their spin measurements will be unreliable.

An alternative approach is to predict halo formation histories from some other approach (e.g. those based on extended Press-Schechter theory), and assign spins to halos in those merger trees in some way. While the distribution of spins is known from N-body simulations (and is largely independent of mass and redshift), and the particle noise present in it can be ``deconvolved'' \citep{benson_constraining_2017}, assigning spins at random from this distribution is not a good approach as we expect spin to be correlated over some timescale at least of order the dynamical time of the halo, and possibly much longer. \cite{cole_hierarchical_2000} attempted to overcome this problem by drawing a spin at random from the measured distribution, but then assuming that this spin remained unchanged until a halo had grown in mass by a factor of 2, at which point a new spin was randomly drawn. This approach ensures correlation in spin across time, but is not well-motivated and its correlation structure has not been tested. \citeauthor{vitvitska_origin_2002}~(\citeyear{vitvitska_origin_2002}; see also \citealt{benson_galaxy_2010}) proposed an alternative model based on the orbital angular momenta of merging halos \cite[an idea also supported by the work of][]{bailin_internal_2005}. Briefly, the angular momentum of any halo is tracked by following the contribution of spin and orbital angular momenta from each halo which merges with it\footnote{The physical origin of angular momentum remains the same as those merging halos gain their orbital angular momentum from large-scale tidal fields.}. \cite{vitvitska_origin_2002} show that this model can reproduce the measured distribution of spin parameters, its independence on mass and redshift, and the enhancement in spin for halos with recent major mergers.

In this work, we develop their model further, by using up to date (and accurately calibrated) models for merger tree construction, halo concentrations, and distributions of orbital parameters for merging halos. We also consider the effects of unresolved accretion, and allow for the possibility of correlation between the angular momenta of infalling satellites and their host halo. This model is then calibrated to the distribution of spins measured in N-body simulations (accounting for particle noise), and the correlation structure of the calibrated model is explored.

\section{Methods}

In this work we make use of two different definitions of spin parameter. The \cite{vitvitska_origin_2002} model directly predicts halo angular momenta, making it simple to compute the corresponding spin under either definition. When constraining the model to match the distribution of spin parameters measured in the Millennium Simulation by \cite{bett_spin_2007} (see \S\ref{sec:constrain}) we utilize the \cite{peebles_origin_1969} definition of spin, as was employed by \cite{bett_spin_2007}:
\begin{equation}
 \boldsymbol{\lambda}_\mathrm{P} = \frac{\boldsymbol{J} |E|^{1/2}}{{\rm G} M^{5/2}},
 \label{eq:spin}
\end{equation}
where $J$ is the magnitude of the halo's angular momentum, $E$ is the energy of the halo (consisting of both gravitational potential and kinetic energy), and $M$ is the halo mass. When examining correlations in spin across time (see \S\ref{sec:correlation}) we utilize the \cite{bullock_universal_2001} definition of spin parameter:
\begin{equation}
\boldsymbol{\lambda}_\mathrm{B} = \boldsymbol{J}/\sqrt{2} M V r,
 \label{eq:spinBullock}
\end{equation}
where $V^2={\rm G}M/r$ with $r$ being the virial radius of the halo. This form is straightforward to compute from data available in the Millennium Simulation database (which does not directly provide halo energies; \citealt{lemson_halo_2006}). As in the above two equations, we will use subscripts P and B to distinguish spins computed using the Peebles and Bullock definitions respectively.

The energy, $E$, of a halo (needed to compute the spin parameter under the \cite{peebles_origin_1969} definition) depends on the density profile of the dark matter halo. Throughout this work we assume NFW \citep{navarro_universal_1997} density profiles, and compute their concentrations using the method of \cite{ludlow_mass-concentration-redshift_2016} as specifically implemented by \cite{benson_halo_2019}.

\subsection{Model for halo spin}

Our model closely follows that of \cite{vitvitska_origin_2002}. Specifically, we begin by building a merger tree using the algorithm of \cite{parkinson_generating_2008} with parameters taken from the posterior distribution found by \cite{benson_halo_2019}. While \cite{vitvitska_origin_2002} used a fixed mass resolution when building their trees we instead adopt a resolution, $M_\mathrm{res}$, which scales with the mass, $M_0$, of the $z=0$ halo in each tree---specifically, we set $M_\mathrm{res} = 10^{-3} M_0$. This choice allows the most massive halos to be processed much more rapidly (as is necessary to facilitate the \MCMC\ simulation described in \S\ref{sec:constrain}), while ensuring that all $z=0$ halos are sufficiently well resolved to have robustly determined spin parameters. We have checked that increasing the resolution (e.g. to $M_\mathrm{res} = 10^{-4} M_0$) makes no significant difference to our results; we find that the distribution of spin parameters for $\lambda_\mathrm{P}$ shifts by less than $0.02$~dex for $\lambda_\mathrm{P} < 0.1$ relative to the $M_\mathrm{res} = 10^{-3} M_0$ case, and shifts by less than $0.05$~dex for $\lambda_\mathrm{P} < 0.2$. We then visit each halo in the tree in a depth-first manner (i.e. visiting all progenitors of a halo before visiting that halo itself) and compute a spin for that halo as follows:

\begin{itemize}
\item \emph{Halos with progenitors:} For halos with one or more progenitors, each such progenitor will already have a spin parameter (and, therefore, an internal angular momentum) assigned. For non-primary progenitors (i.e. those which will merge with the primary progenitor and become subhalos) we assign orbital parameters at the point of merging as described in \S\ref{sec:orbits}. We then simply sum the spin and orbital angular momenta in the centre of mass frame of the primary-secondary progenitor system. We allow for the orbital angular momentum to be divided by a factor $(1+M_2/M_1)^{1-\epsilon}$, where $M_1$ and $M_2$ are the masses of the primary and secondary progenitors respectively, to allow for the possibility that orbital angular momentum is not conserved in the merger\footnote{Since mass can be lost from halos during major mergers \protect\citep{lee_tidal_2018} angular momentum may also be lost.}. We treat $\epsilon$ as a parameter of the model to be determined.

Since our merger trees have a finite mass resolution some accretion onto halos will be unresolved. With higher resolution this unresolved accretion would break up into low mass progenitor halos which would contribute to the angular momentum of each halo. To account for angular momentum contributed by these ``sub-resolution'' halos we assume that such unresolved accretion contributes angular momentum at the mean rate found by averaging over the orbital parameter distribution as described in \S\ref{sec:orbits}. The contribution made to the angular momentum by sub-resolution halos will depend on the mass of those halos. This mass dependence arises from the $(1+M_2/M_1)^{1-\epsilon}$ factor above, where $M_2$ is the mass of the sub-resolution halo. We therefore average the mean angular momentum (see \S\protect\ref{sec:orbits}) of sub-resolution halos, including this factor, over the mass function of sub-resolution halos. We assume a mass function slope of $\alpha=-1.9$ \protect\citep{springel_aquarius_2008} for sub-resolution halos. The resulting angular momentum due to accretion of sub-resolution halos is then
\begin{equation}
\boldsymbol{J}_\mathrm{unresolved} = {2-\alpha \over \mu^{2-\alpha}} B(\mu/[1+\mu];2-\alpha,\epsilon-2+\alpha) M_\mathrm{unresolved} \langle \boldsymbol{j} \rangle,
\label{eq:unresolved}
\end{equation}
where $B(x;a,b)$ is the incomplete beta function, $M_\mathrm{unresolved}$ is the mass accreted in sub-resolution halos, and $\langle \boldsymbol{j} \rangle$ is the mean specific orbital angular momentum of sub-resolution halos (see \S\ref{sec:orbits}). In Appendix~\ref{sec:resolution} we show that our results are well-converged with respect to resolution.

\item \emph{Progenitorless halos:} For halos with no progenitor we assign a spin parameter by drawing at random from a distribution---specifically we use the functional form of \cite{bett_spin_2007} with parameters taken from the posterior distribution found by \cite{benson_constraining_2017}. 
\end{itemize}

Applying this procedure to a merger tree results in a determination of the internal angular momentum of each halo---these can be converted to spins following the usual definition (equations \ref{eq:spin} and \ref{eq:spinBullock}).

\subsubsection{Orbital parameters of progenitor halos}\label{sec:orbits}

Orbital parameters of merging halos are drawn from the distributions reported by \cite{jiang_orbital_2015}, including the dependence on primary halo mass and secondary/primary halo mass ratio. The \cite{jiang_orbital_2015} results give the radial and tangential velocities of each merging secondary halo as it crosses the virial radius of the primary halo\footnote{We note that \protect\cite{jiang_orbital_2015} adopt a definition of virial radius corresponding to the radius enclosing a mean interior density of 200 times the critical density. As this differs from that definition used in building our merger trees---which assume a spherical collapse model for defining the virial radius---we propagate the orbital velocities drawn from the \cite{jiang_orbital_2015} distribution to our preferred definition of virial radius assuming that energy and angular momentum are conserved along the orbit.}. This specifies three of the six phase-space coordinates of the secondary halo. Previous works \citep{vitvitska_origin_2002,benson_galaxy_2010} have fixed the remaining three parameters by assuming that merging secondaries are distributed uniformly over the virial sphere of their primary, and that tangential velocities are isotropically distributed\footnote{\protect\cite{vitvitska_origin_2002} examined the orbital parameters of infalling subhalos in cosmological N-body simulations but found no significant correlations. However, as they noted, given the statistical power of their sample such correlations could still be present at the level of 10--20\%.}. We move beyond this assumption and allow for the possibility of some correlation in the orbital parameters of secondaries. Specifically, we allow for a correlation between the orbital angular momentum, $\boldsymbol{J}_\mathrm{orb}$, of the secondary, and the vector spin\footnote{That is, a vector with magnitude equal to the spin parameter, and direction coincident with the internal angular momentum vector of the halo.}, $\boldsymbol{\lambda}_\mathrm{P}$, of the primary, such that the angle $\theta$ between these two vectors is distributed as:
\begin{equation}
P(\cos\theta) = \frac{1}{2} \left(1 + \alpha |\boldsymbol{\lambda}_\mathrm{P}| \cos \theta \right),
\label{eq:correlatedOrbits}
\end{equation}
where $\alpha$ is a parameter which controls the strength of the correlation\footnote{For $\alpha |\boldsymbol{\lambda}_\mathrm{P}| > 1$ this function becomes negative, making it an invalid distribution. In practice we find that this does not occur for values of $\alpha$ required to fit N-body simulations---see Appendix~\protect\ref{sec:posterior}.}, and which we treat as a parameter of the model to be determined. Recently, \protect\cite{an_living_2020} have shown a strong anisotropy in the spin-orbit alignment of merging pairs of halos of comparable masses (mass ratios of $3:1$ or less), with a distribution function similar in form to eqn.~(\protect\ref{eq:correlatedOrbits}).

To sample orbits from this distribution we first draw orbital parameters from the distribution of \cite{jiang_orbital_2015}, and choose the remaining phase space coordinates assuming isotropically distributed infall on the virial sphere and isotropically distributed tangential velocities. We compute the resulting angular momentum vector, and from it determine $\cos\theta$. We then use rejection sampling, by accepting the orbit with probability
\begin{equation}
 P(\cos\theta) = {1 + \alpha |\boldsymbol{\lambda}_\mathrm{P}| \cos \theta \over 1 + \alpha |\boldsymbol{\lambda}_\mathrm{P}|},
\end{equation}
to produce a distribution consistent with equation~(\ref{eq:correlatedOrbits}).

Since eqn.~(\ref{eq:correlatedOrbits}) depends only on the angle $\theta$ it is clear that the mean specific angular momentum of infalling halos must be aligned (or anti-aligned) with $\boldsymbol{\lambda}_\mathrm{P}$. For any given infalling halo the magnitude of its specific angular momentum along the direction of $\boldsymbol{\lambda}_\mathrm{P}$ will be $j = r_\mathrm{v} v_\phi \cos\theta$. Given the distribution in equation~(\ref{eq:correlatedOrbits}) the mean specific angular momentum of infalling halos is therefore
\begin{equation}
 \langle \boldsymbol{j} \rangle = \int_0^\infty \mathrm{d}v_\phi \int_{-1}^{+1} \mathrm{d}(\cos\theta) r_\mathrm{v} v_\phi f(v_\phi) \cos\theta P(\cos\theta),
\end{equation}
where $f(v_\phi)$ is the distribution of the tangential component of orbital velocity of infalling halos. This integral is easily evalutated to give
\begin{equation}
 \langle \boldsymbol{j} \rangle = {\alpha \over 3} r_\mathrm{v} \boldsymbol{\lambda}_\mathrm{P} \langle v_\phi \rangle,
\end{equation}
where $\langle v_\phi \rangle = \int_0^\infty \mathrm{d}v_\phi v_\phi f(v_\phi)$ is the mean tangential velocity for infalling halos which we computed using the distribution of \protect\cite{jiang_orbital_2015}.Given the distribution in equation~(\ref{eq:correlatedOrbits}) the mean specific angular momentum of infalling halos is $\langle \boldsymbol{j} \rangle = \alpha \boldsymbol{\lambda}_\mathrm{P} \langle v_\theta \rangle / 3$ where $\langle v_\theta \rangle$ is the mean tangential orbital velocity from the distribution of Jiang et al. (2015).

\subsection{Constraining parameters of the model}\label{sec:constrain}

The process described in the preceding sub-section is repeated for a large number of merger trees, using cosmological parameters and a power spectrum matched to the Millennium Simulation \citep{springel_simulations_2005}, $z=0$ halo masses drawn from a \cite{sheth_ellipsoidal_2001} mass function (with parameters given by \citealt{benson_halo_2019}) and spanning the range $3.53\times10^{11}\mathrm{M}_\odot$ to $1.00\times10^{15}\mathrm{M}_\odot$ to match the selection used by \cite{bett_spin_2007}. \cite{bett_spin_2007} used a ``quasi-equilibrium'' criterion, based on the virial ratio $2T/U+1$ (with $T$ and $U$ being the kinetic and potential energies of the halo respectively) to remove halos from their sample which were far from virial equilibrium. Since we can not compute the dynamical evolution of the virial ratio of halos in our model we instead remove halos which are likely to be unrelaxed based on a major merger criterion. Specifically, we exclude from our sample any $z=0$ halo which experience a merger with mass ratio $M_2/M_1>f_\mathrm{major}$ more recently than a look-back time of $t_\mathrm{major}$. We treat $f_\mathrm{major}$ and $t_\mathrm{major}$ as nuisance parameters when constraining the parameters of our model.

The spin of each remaining $z=0$ halo computed in this way is convolved with the distribution function describing the effects of particle noise on measurement of N-body halo spins using the model of \cite{benson_constraining_2017}. The quasi-equilibrium selection criterion that \cite{bett_spin_2007} imposed on their halo sample was designed to remove any halos which are far from virial equilibrium. The model of \cite{benson_constraining_2017} contains a log-normal component which models deviations of the mass and energy of an N-body halo from their true values due to particle noise. Any halos which experience a very large deviation in energy because of particle noise would be excluded from the \cite{bett_spin_2007} sample by their quasi-equilibrium criterion. Therefore, when convolving halo spins with the \cite{benson_constraining_2017} distribution function we truncate the log-normal component beyond values that are a factor $R$ above or below the mean. In this way we avoid populating the tails of the distribution which would correspond to halos excluded by the quasi-equilibrium criterion. We allow some freedom in the factor $R$ as will be discussed below.

The results are summed over all merger trees to give the final distribution of spin parameters as would be measured in the Millennium Simulation using the approach of \cite{bett_spin_2007}. We then compute the likelihood of N-body results of \cite{bett_spin_2007} given our model using:
\begin{equation}
 \log \mathcal{L} = -\frac{1}{2} \Delta \mathbfss{C}^{-1} \Delta^{\rm T},
\end{equation}
where $\Delta$ is a vector of differences between the spin distribution of \cite{bett_spin_2007} and that predicted by our model. The covariance matrix, $\mathbfss{C}=\mathbfss{C}_\mathrm{N-body}+\mathbfss{C}_\mathrm{model}$, where the covariance matrix of the N-body data, $\mathbfss{C}_\mathrm{N-body}$, is assumed to be diagonal and equal to the Poisson variance in each bin, and the covariance matrix of our model calculation, $\mathbfss{C}_\mathrm{model}$, is computed following the approach of \cite{benson_building_2014} accounting for the correlations introduced between bins by the process of convolving with the particle noise distribution.

The parameters of our spin model are then calibrated by running a \MCMC\ simulation, following the approach of \cite{benson_mass_2017} in detail, including utilizing the same \MCMC\ algorithm and convergence criteria. Briefly, we perform a differential evolution \MCMC\ simulation \citep{terr_braak_markov_2006} using 128 parallel chains. At each step of the simulation a proposed state, $S_i^\prime$, for each chain, $i$, is constructed by selecting at random (without replacement) two other chains, $m$ and $n$, and finding
\begin{equation}
 S_i^\prime = S_i + \gamma (S_m - S_n) + \epsilon,
\end{equation}
where $\gamma$ is a parameter chosen to keep the acceptance rate of proposed states sufficiently high, and $\epsilon$ is a random vector each component of which is drawn from a Cauchy distribution with median zero and width parameter set equal to $10^{-9}$ of the current range of parameter values spanned by the ensemble of chains to ensure that the chains are positively recurrent. For a multivariate normal likelihood function in $N$ dimensions the optimal value of $\gamma$ is $\gamma_0=2.38/\sqrt{N}$ \citep{terr_braak_markov_2006}. We use this as our initial value of $\gamma$, but adjust $\gamma$ adaptively as the simulation progresses to maintain a reasonable acceptance rate. The proposed state is accepted with probability $P$ where
\begin{equation}
 P = \left\{ \begin{array}{ll} 1 & \hbox{ if } \mathcal{L}(S_i^\prime) > \mathcal{L}(S_i), \\ \mathcal{L}(S_i^\prime)/\mathcal{L}(S_i) & \hbox{ otherwise,} \end{array} \right.
\end{equation}
and where $\mathcal{L}$ is the likelihood function.

The simulation is allowed to progress until the chains have converged on the posterior distribution as judged by the Gelman-Rubin statistic, $\hat{R}$ \citep{gelman_a._inference_1992}, after outlier chains (identified using the Grubb's outlier test \citep{grubbs_procedures_1969,stefansky_rejecting_1972} with significance level $\alpha=0.05$) have been discarded. Specifically, we declare convergence when $\hat{R}=1.2$ in the parameters of interest, $\epsilon$ and $\alpha$.

The Gelman-Rubin convergence measure relies on the chains be initialized in an over dispersed state. The state of each chain is therefore initialized by constructing 128-point unit Latin hypercubes. We generate 100 such cubes and find the cube which maximizes the minimum ($\ell^2$-norm) distance between any two points in the hypercube. Each point in this hypercube realization is used as the initial state for a chain by associating $C_i=L_i$ where $L_i$ is the $i^{\rm th}$ coordinate of the point in the hypercube, and $C_i$ is the cumulative probability distribution of the prior on parameter $i$. The parameter values are then simply found by inverting their cumulative distributions. 

We allow the following parameters to vary in this \MCMC\ simulation:
\begin{itemize}
 \item $(A,a,p)$ in the \cite{sheth_ellipsoidal_2001} mass function, $(G_0,\gamma_1,\gamma_2)$ in the halo merger rate model of \cite{parkinson_generating_2008}, and $(f,C)$ in the halo concentration model of \cite{ludlow_mass-concentration-redshift_2016} all as defined by \cite{benson_halo_2019}, with a multivariate normal prior matched to the posterior distribution found by \cite{benson_halo_2019}.
 \item $(\lambda_{P,0},\alpha)$ in the spin parameter distribution fitting function\footnote{This distribution is used to sample spin parameters for halos with no progenitors.} of \cite{bett_spin_2007}, with a multivariate normal prior matched to the posterior distribution found by \cite{benson_constraining_2017}.
 \item $(b,\gamma,\sigma,\mu)$ in each primary halo mass, and secondary-to-primary mass ratio range in the fitting function for orbital parameters of subhalos of \cite{jiang_orbital_2015}, with normal priors with means and variances derived from the best-fit values and errors reported by \cite{jiang_orbital_2015}.
 \item The major merger mass ratio, $f_\mathrm{major}$, and time, $t_\mathrm{major}$, used to exclude halos with recent major mergers, with normal priors with (mean, variance) of $(0.20,0.01)$ and $(0.50,0.09)$ respectively. The prior for $f_\mathrm{major}$ is motivated by the fact that \cite{bett_spin_2007} excluded halos with $|2T/U+1|<Q$ with $Q=0.5$. Perturbations of this magnitude to virial equilibrium should be expected to require mergers of mass ratio roughly comparable to $Q$. The timescale for exclude halos with recent major mergers is motivated by the study of \cite{drakos_major_2018} who find that virial equilibrium is reestablished within around 2~Gyr after first passage for binary (i.e. equal mass) mergers of typical $z=0$ halos, we therefore expect a revirialization timescale shorter than this for major but non-binary mergers.
\item The factor $R$ at which the log-normal component of the \cite{benson_constraining_2017} particle noise distribution function is truncated to mimic the effects of the \cite{bett_spin_2007} quasi-equilibrium selection criterion. We expect this factor to be $R \approx 1 + Q$, but allow some freedom by adopting a uniform prior in the range $1.4$ to $2.0$.
 \item $\epsilon$ in our model for the angular momentum retained by halos during mergers, with a uniform prior between $0$ and $3$. A value of $\epsilon=1$ indicates no loss of angular momentum during mergers. Since major mergers can lead to mass loss, and therefore angular momentum loss, we may expect $\epsilon > 1$. \cite{lee_tidal_2018} find mass loss at the level of 10\% following major ($M_2/M_1 >0.3$) mergers. If 10\% of angular momentum was lost in such cases it would imply $\epsilon\approx 1.4$. However, since mass is lost from the outer regions of halos it is likely that the specific angular momentum of the lost material is higher than average. The upper limit of our prior is therefore chosen be sufficiently high to allow for the specific angular momentum to be enhanced by a factor of around $1.5$ above the average. While the above arguments regarding angular momentum loss suggest $\epsilon > 1$ we allow our prior to extend to $\epsilon=0$---our model is idealized in several respects (e.g. it assumes spherical halos) so we allow for the possibility that a better match to the N-body results may be obtained with $\epsilon < 1$.
 \item $\alpha$ in our model for correlated infall orbits with a prior that is uniform in $\alpha$ between $\alpha=0$ and $20$. While \cite{vitvitska_origin_2002} measured no correlations between orbital parameters of infalling satellites and their host halos, the small size of their sample left the possibility of a correlation at the 10--20\% level. In our model for correlated orbits this would correspond to $\alpha$ in the range 2--3 for typical spin parameters. \protect\cite{an_living_2020} show results for approximately equal-mass mergers of halos which correspond to a much larger $\alpha \sim 20$. We allow a broad range of $\alpha$ which encompass these expectations.
\end{itemize}
This gives a total of $50$ parameters, although most of them are well-constrained by previous analyses.

\section{Results}

Our \MCMC\ simulation reaches convergence (as judged by the Gelman-Rubin statistic---see \S\ref{sec:constrain}) after 2,090 steps. We discard these initial burn-in steps then allow our \MCMC\ simulation to run for a further 2,450 steps. The posterior distribution over the model parameters is then determined from those post-convergence steps. The correlation length in our chains is around 26 steps. Therefore, with 128 chains we have approximately 12,000 independent draws from the posterior distribution. We find that the parameters of our spin model are constrained to be $ \epsilon$ $=(2.40^{+1.20}_{-0.80}) \times 10^{-1}$ and $ \alpha$ $=(6.6^{+11.3}_{-4.5}) \times 10^{-1}$. The parameters controlling removal of halos with recent major mergers are constrained to be $ f_\mathrm{major}$ $=(1.90^{+0.71}_{-1.06}) \times 10^{-1}$, $ t_\mathrm{major}$ $=(2.8^{+4.3}_{-1.4}) \times 10^{-1}$~Gyr, and $ R$ $=1.420^{+0.065}_{-0.022}$. The remaining nuisance parameters have posterior distributions largely consistent with their priors.

Interestingly the posterior distribution of $\epsilon$ is found to be almost entirely constrained to the $\epsilon < 1$ region, contradicting our expectation that angular momentum loss in major mergers would lead to $\epsilon > 1$. The posterior also favours a non-zero $\alpha \approx 0.9$ indicating that some correlation between subhalo orbits is preferred, although not as strong as that reported by \protect\cite{an_living_2020} for approximately equal-mass mergers. We find (see Appendix~\ref{sec:posterior}) that $\epsilon$ and $\alpha$ are strongly correlated. We will comment further on this correlation and its implications in \S\ref{sec:correlation}.

\subsection{Spin distribution}

\begin{figure}
 \includegraphics[width=85mm]{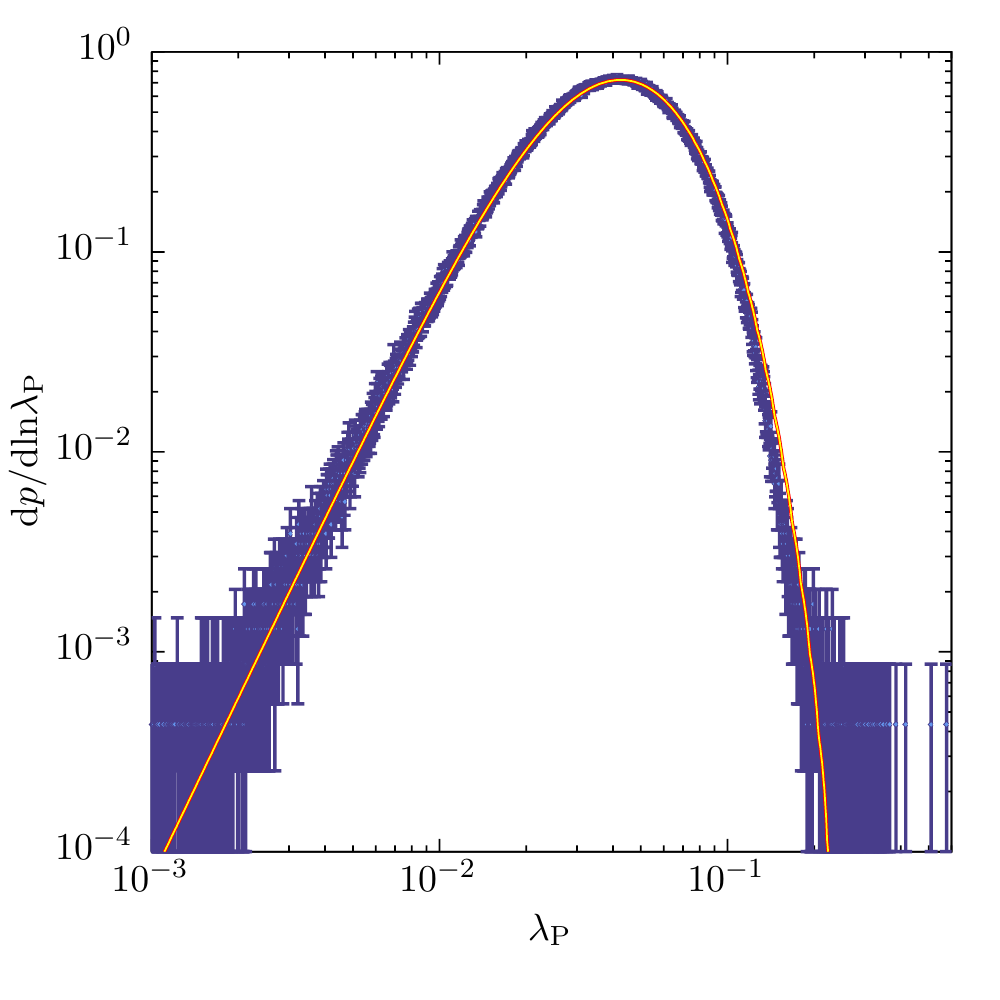}
 \caption{The distribution of spin parameters, $\lambda_\mathrm{P}$, for relaxed halos at $z=0$. The blue points with error bars show the results of \protect\cite{bett_spin_2007}, while the yellow line shows the results from our model for the maximum posterior probability model after convolution with the particle noise distribution predicted by the model of \protect\cite{benson_constraining_2017}.}
 \label{fig:spinDistribution}
\end{figure}

Using the maximum posterior probability model found by our \MCMC\ simulation we compute the $z=0$ distribution of (Peebles) spin parameters for the halo mass range used by \cite{bett_spin_2007}, convolve them with the particle noise distribution of \cite{benson_constraining_2017} and compare them with the distribution measured by \cite{bett_spin_2007} for N-body halos. The results are shown in Figure~\ref{fig:spinDistribution}.

It is apparent that our model matches the N-body results extremely well. In particular, the position of the peak in the distribution, the width of the distribution, and the low-$\lambda_\mathrm{P}$ slope are all almost perfectly reproduced. At high-$\lambda_\mathrm{P}$ our model slightly overpredicts the distribution function, although this is a regime where the details of how unrelaxed halos are rejected from the sample and the specifics of the particle noise distribution have the greatest effect. Less than 10\% of halos have spins in this regime.

\subsection{Correlations}\label{sec:correlation}

To examine the correlation structure of the spin of a halo across time we first extract a set of merger trees with $z=0$ halo masses in the mass range $3.53\times10^{13}$ to $10^{15}\mathrm{M}_\odot$ from the Millennium Simulation database. Halos in this mass range in the Millennium Simulation contain at least 30,000 particles, and so their spins are only mildly affected by particle noise. We compute spins for all halos in these trees under the \cite{bullock_universal_2001} definition, and then extract the time series of spin along the main branch of the merger tree back to early times. We then measure the correlation of spin parameter magnitude of $z=0$ halos with that of their progenitor at early times, and also measure the correlation of $\cos \theta$, where $\theta$ is the angle between the spin vectors of the $z=0$ halo and its earlier progenitor.

At each look-back time the correlation functions are computed by averaging over all $z=0$ halos in our sample which have a resolved progenitor at that time. Therefore, the number of $z=0$ halos averaged over at each look-back time will decrease as look-back time increases due to the fact that there is a maximal look-back time for each $z=0$ halo at which a resolved progenitor halo exists. This will have two effects on the correlation function:
\begin{enumerate}
\item The correlation function will become noisier at large look-back times due to the smaller number of $z=0$ halos contributing (noise is also contributed at large look-back times by the fact that the progenitor halos consist of ever fewer particles, due to their decreasing masses, making their spins less well-determined);

\item Some bias may be introduced if, for example, the auto-correlation function, $\mathrm{Corr}(|\boldsymbol{\lambda}_\mathrm{P}|)$ (the correlation between the magnitude of spin of the $z=0$ halo and its progenitor at some earlier time), is itself correlated with halo formation history (which, in our model, it of course is) since trees which formed earlier will have larger maximal look-back times.
\end{enumerate}

For comparison, we construct a sample of merger trees using the model described in this work using the maximum posterior probability model found by our \MCMC\ simulation, spanning the same range of masses and in a cosmology matched to that of the Millennium Simulation. These merger trees are constructed with the mass halo mass resolution as those taken from the Millennium Simulation. As such, our merger trees have a distribution of maximal look-back times which is consistent with those from the Millennium simulation (as expected given that we use a merger tree building algorithm which has been calibrated to N-body simulation progenitor mass functions).

After computing angular momenta of each halo in these trees we use the model of \cite{benson_halo_2019} to add noise to the angular momenta and masses of each halo to represent the effects of particle noise---while this is generally a small effect for halos containing 30,000 particles or more as in this sample we nevertheless account for the effects of this noise\footnote{The noise becomes more significant for progenitor halos at early times which have significantly lower masses.}. Our merger trees (to the extent that our model is a valid description of the physics governing halo spins) should have the same statistical and systematic uncertainties as the N-body data to which we compare. Finally, we compute the spin parameters of these halos under the \cite{bullock_universal_2001} definition and compute correlations in spin magnitude and direction in the same way as for the Millennium Simulation halos.

\begin{figure}
 \includegraphics[width=85mm]{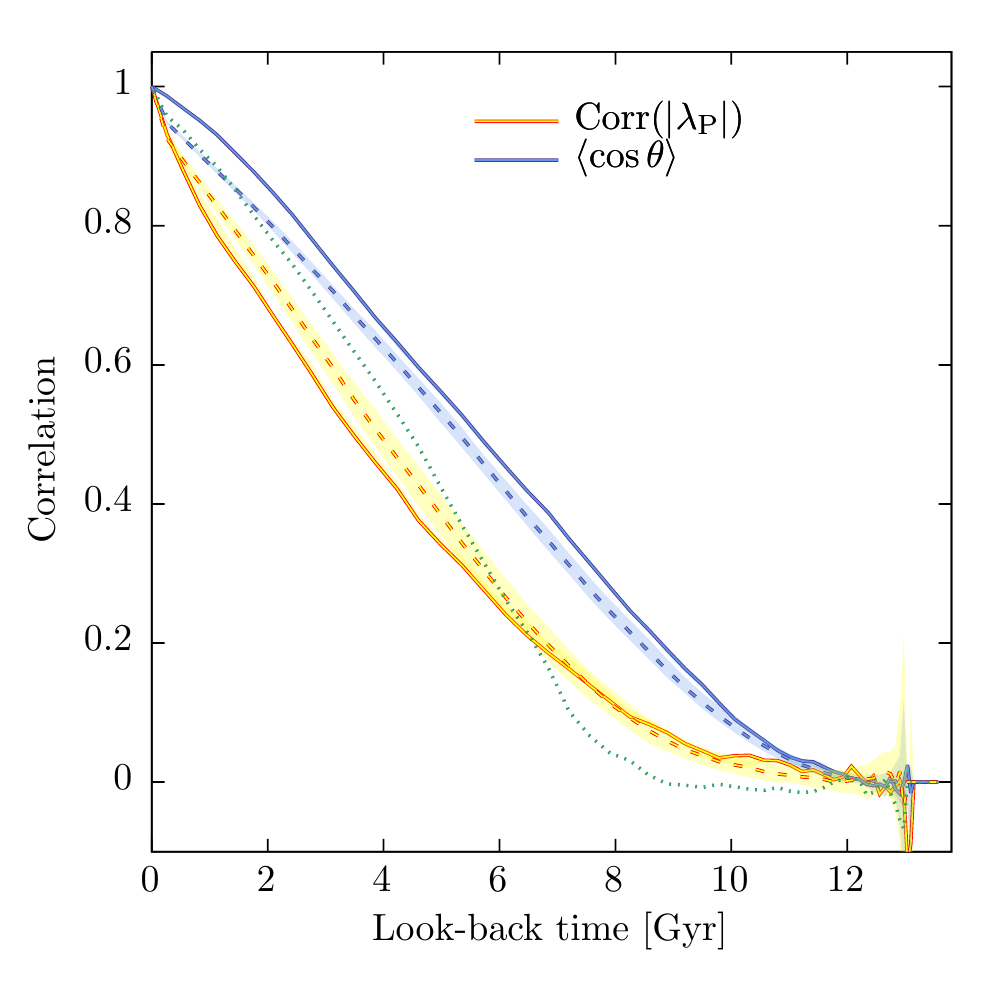}
 \caption{Correlation functions of spin magnitude (yellow lines) and angle (blue lines) from the Millennium simulation (solid lines), and from our model (dashed lines). For the Millennium simulation the line shows the result averaged over all halos, with the shaded region indicating the 10--90\% inter-percentile region over all halos. For our model the line indicates the median from the posterior distribution, with the shaded region indicating the 10--90\% inter-percentile region of the posterior distribution. The dotted green line shows the correlation function for spin magnitude in the Galform model \protect\citep{cole_hierarchical_2000} in which spins are selected at random, and updated each time a halo doubles its mass.}
 \label{fig:autocorrelation}
\end{figure}

Figure~\ref{fig:autocorrelation} shows the resulting correlation functions for the Millennium Simulation \citep{springel_simulations_2005} and for our model. Considering first the correlation in the magnitude of spin (yellow lines), our model (dashed line) matches that measured in the Millennium Simulation (solid line) very well across the full range of look-back times, with only a slight over-estimation of correlation at look back times less than 6~Gyr. The dotted green line in Figure~\ref{fig:autocorrelation} shows the Galform model \citep{cole_hierarchical_2000} in which spins are selected at random from a distribution, and updated each time a halo doubles its mass. It can be seen that this does not produce the correct correlation structure as measured from the Millennium Simulation, over-predicting the correlation up to look-back times of 6~Gyr, beyond which it underpredicts (at a level comparable to that underprediction found using the model in this work). We will briefly explore the consequences of these differences in \S\ref{sec:discussion}.

In terms of the angle between the $z=0$ spin vector and that at earlier times (blue lines), our model (dashed line) decorrelates somewhat faster than the N-body halos (solid line) with look-back time, but reproduces the overall trend, and matches the N-body correlation quite well for larger look-back times. This suggests that our model for the correlations between infalling subhalo orbits does not capture the true correlations sufficiently well.

Recently, \cite{morinaga_impact_2019} examined the filamentary nature of subhalo accretion, finding correlations between the strength of filamentary accretion and halo shape and orientation. This clearly demonstrates that the correlations arising from the filamentary nature of accretion can have measurable effects on halo properties. They find that the angular momentum vectors of halos tend to be aligned perpendicularly with the direction of filaments \cite[see also][]{libeskind_cosmic_2012}.

Our model for correlated infall orbits was chosen to be simple and empirical, and, for example, does not explicitly account for the filamentary nature of subhalo accretion. The results shown in Figure~\ref{fig:autocorrelation} suggest that this model should be improved, perhaps by studying these correlations directly in N-body simulations and developing a model to describe them. Furthermore, we find (see Appendix~\ref{sec:posterior}) that the parameter describing correlated infall orbits is strongly correlated with the parameter $\epsilon$ which parameterizes the non-conservation of angular momentum in major mergers. We find $\epsilon < 1$ which implies that major mergers contribute more angular momentum than expected. An improved model for the correlated nature of infall orbits may therefore also change the inference for $\epsilon$---a model which gave stronger correlation between the angular momentum vectors of infalling satellites and the spin of the host halo would presumably allow $\epsilon$ to increase, as the mean angular momentum per merger would then be larger. Alternatively, the inference of $\epsilon < 1$ may indicate that our model is too simplistic to capture the details of major merger events---examination of the angular momentum content of well-resolved cosmological N-body halos undergoing major mergers may shed light on this aspect of the model.

\subsection{Further Tests}

Utilizing the same sample of merger trees as in \S\ref{sec:correlation} we can construct further tests of how well our model matches measurements related to halo spin in N-body simulations. In the following we will explore how the spin depends on halo formation history, and examine the distribution of ``spin-flips'' (i.e. large changes in the direction of the spin vector).

\subsubsection{Dependence on Halo Formation Time}

We first consider how the median halo spin correlates with halo formation time, $t_\mathrm{f}$. We adopt a conventional definition of formation time as that time at which the primary progenitor of a $z=0$ halo first reaches 50\% of the mass of that $z=0$ halo. We compute formation times using the merger trees our model (i.e. built using the \protect\cite{parkinson_generating_2008} algorithm), and from the Millennium Simulation, using the same sample of $z=0$ halos as in \S\ref{sec:correlation}. We then compute the median spin parameter in bins of formation time.

\begin{figure}
 \includegraphics[width=85mm]{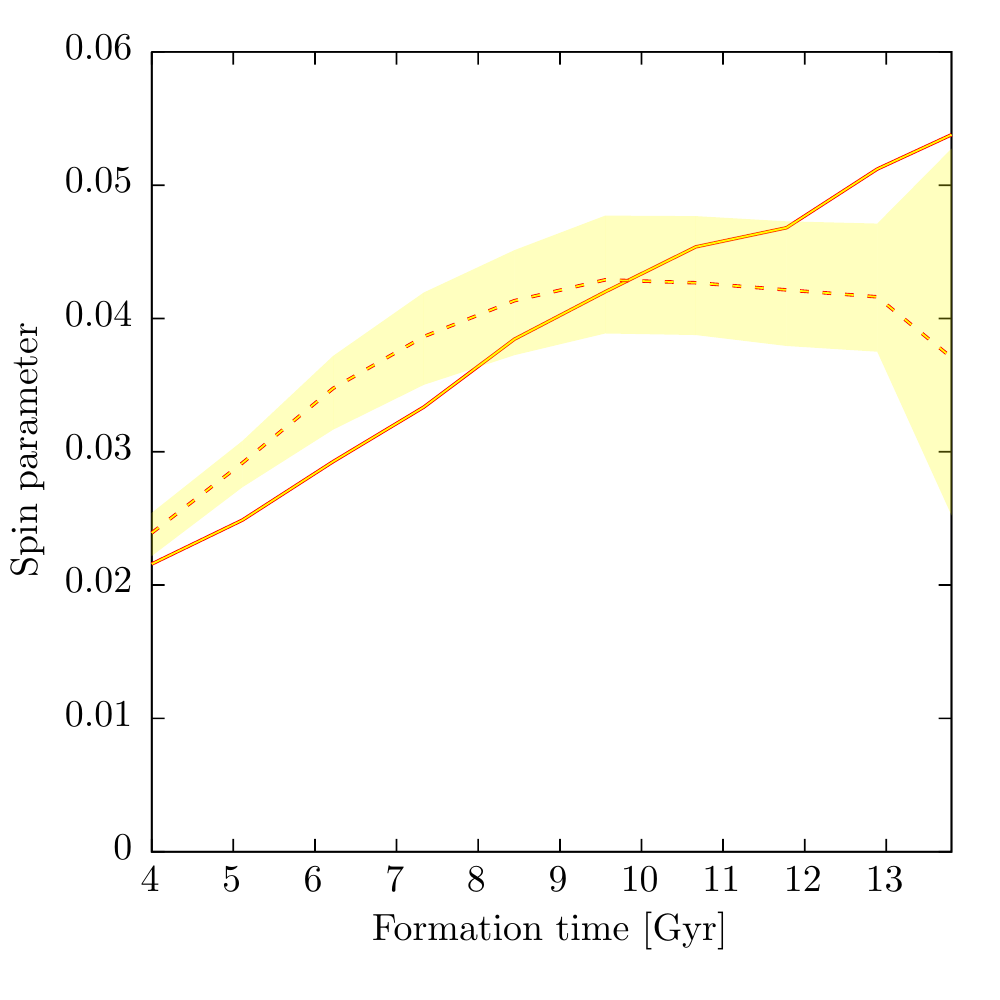}
 \caption{The median (Bullock) spin parameter for $z=0$ halos as a function of halo formation time---defined as the time at which the primary progenitor of the $z=0$ halo first reaches 50\% of the $z=0$ halo mass. The solid line indicates results from the Millennium Simulation, while the dashed line and shaded region indicate the median and 10\% and 90\% percentiles of the posterior distribution from our model.}
 \label{fig:formationTime}
\end{figure}

Figure~\ref{fig:formationTime} shows the results of this analysis. There is a clear trend for the median spin to increase as formation time increases. Note that, in this analysis, we have not excluded any halos on the basis of recent mergers or departures from virial equilibrium. In the case of the Millennium Simulation (solid line), the medin spin increases from close to $\bar{\lambda}_\mathrm{B}=0.020$ for halos formed at $t=4$~Gyr, to $\bar{\lambda}_\mathrm{B}=0.054$ for halos formed very recently. Comparing the the posterior predictions from our model, we see that the trend is matched quite well for early-forming ($t_\mathrm{f} <9$~Gyr) halos, but for later-forming halos our model predicts no significant correlation between formation time and median spin.

Halos which formed recently are more likely to have recent major merger events than those which formed early. We therefore interpret the trend found in the N-body simulation as indicating that, on average, major mergers temporarily increase the halo spin, while extended periods of slow accretion (consisting of many minor mergers) cause a regression toward the mean and halo spin decreases. Our model fails to produce a sufficiently strong trend of median spin with formation time at late times. This may indicate that our model has too many minor mergers/too few major mergers at late times (a known problem in the \protect\cite{parkinson_generating_2008} merger tree algorithm---see \protect\citeauthor{benson_mass_2017}~(\protect\citeyear{benson_mass_2017}; Fig.~7) for example).

\subsubsection{Distribution of ``Spin Flips''}

We next examine the distribution of ``spin-flips'' \citep{bett_spin_2012}, that is, the distribution of the change in angle of the spin vector over short time intervals. To examine this we measure the quantities:
\begin{equation}
 \Delta \mu (t) = {M(t) - M(t-\tau) \over M(t)},
\end{equation}
and
\begin{equation}
 \cos \theta(t) = {\boldsymbol{J}(t)\cdot\boldsymbol{J}(t-\tau) \over |\boldsymbol{J}(t)| |\boldsymbol{J}(t-\tau)|},
\end{equation}
defined by \protect\cite{bett_spin_2012}, and which measure the fractional change in the mass of a halo, and the change in the angle of the halo's angular momentum vector over a timescale $\tau$. \protect\cite{bett_spin_2012} choose $\tau = 0.5$~Gyr. We choose a value of $\tau = 0.54$~Gyr which is close to that of \protect\cite{bett_spin_2012} and corresponds precisely to an interval between snapshots of the Millennium Simulation and so avoids the need for any interpolation.

\begin{figure}
 \includegraphics[width=85mm]{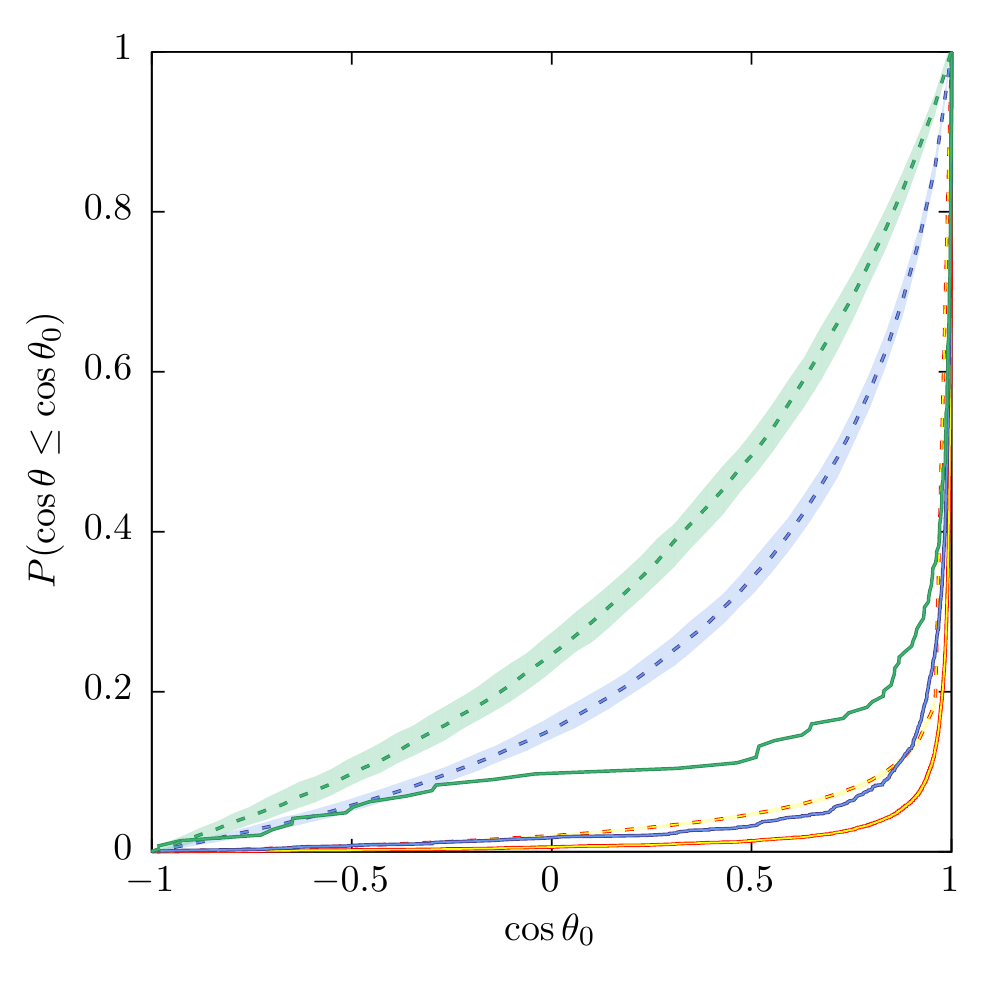}
 \caption{The cumulative distribution of spin-flip angles, $\theta$, for halos at $z=0$ measured over a time interval of $0.54$~Gyr. Solid lines indicate results from the Millennium Simulation, while the dashed lines and shaded regions indicate the median and 10\% and 90\% percentiles of the posterior distribution from our model. The yellow, blue, and green lines show results for mass fraction changes of $\Delta \mu \ge 0$, $0.1$, and $0.3$ respectively.}
 \label{fig:spinFlips}
\end{figure}

In Figure~\protect\ref{fig:spinFlips} we show the cumulative distribution of $\cos \theta$ for our sample of $z=0$ halos, for subsamples with $\Delta \mu \ge 0$, $0.1$, and $0.3$ (yellow, blue, and green lines respectively). For the N-body simulation (solid lines) the distribution is strongly concentrated around $\cos\theta=1$ (i.e. no change in angle) with a long tail to smaller values. For larger $\Delta \mu$ (i.e. halos which have experienced significant recent mass growth) the distribution becomes more extended.

Comparing the results from our model we see that, for the $\Delta \mu > 0$ subsample the distribution of $\cos\theta$ is also strongly concentrated around $\cos\theta=1$, although the tail to lower $\cos\theta$ is heavier. For the $\Delta \mu \ge 0.1$, and $0.3$ subsamples though our model predicts a significantly more extended tail. This difference from the N-body results is closely related to our finding of $\epsilon < 1$, which enhances the angular momentum provided by major mergers, and therefore causes such mergers to result in much larger changes in the direction of the angular momentum vector.

\section{Discussion}\label{sec:discussion}

We have described an updated implementation of the random walk model for halo angular momentum first proposed by \cite{vitvitska_origin_2002}. Our approach utilizes updated algorithms for merger tree construction, more accurate distributions of orbital parameters, correlations between those parameters, and is carefully calibrated to measurements from N-body simulations accounting for the effects of particle noise. The resulting algorithm is implemented within the open-source semi-analytic model {\sc Galacticus}\protect\footnote{\protect\href{github.com/galacticusorg/galacticus}{{\tt github.com/galacticusorg/galacticus}}} and so is available for any one to use. We also provide an overview of our algorithm in the form of a flowchart in Appendix~\protect\ref{app:pseudocode}. The resulting model accurately reproduces the measured distribution of spin parameters from N-body simulations for $\lambda_\mathrm{P} < 0.1$, slightly overpredicting the distribution at larger spins. Achieving this match requires non-zero correlations between the angular momenta of infalling halos and the spin of the halo with which they are merging. Although this can not be tested in our model, it is expected that these correlations derive from the large scale tidal field around the host halo which is coherent over cosmological timescales, and the filamentary nature of subhalo accretion.

Once calibrated to match the distribution of spin parameters in halos, the model can be used to explore correlations in spin parameter magnitude and direction over time in halos. We find that the model accurately matches the correlation measured from N-body simulations in the magnitude of the spin up to a look-back time of 4~Gyr, somewhat underpredicting the correlation at larger look-back times. The agreement with N-body results is significantly improved with respect to algorithms previously used in semi-analytic galaxy formation models \citep{cole_hierarchical_2000}. Correlation in the direction of the spin vector is somewhat underestimated with respect to measurements from N-body simulations, suggesting that our simple model for correlations in the orbital parameters of infalling subhalos is insufficient, and should be improved (perhaps by understanding developed from studies of N-body simulations). This point is further emphasized by considering the distribution of ``spin flips'' \protect\citep{bett_spin_2012,bett_spin_2016} in which our model shows much larger changes in the direction of the spin vector as a result of large mass accretion events than is found in N-body simulations. These and similar statistics will therefore provide a useful test of any improvements to our model resulting from further studies of correlations in the orbital parameters of infalling subhalos. We also find that our model qualitatively reproduces trends between spin parameter and halo formation time, although quantitative differences remain.

The random walk model described here allows for spins to be assigned to dark matter halos in merger trees in a manner that is internally consistent with their formation histories. Since halo concentrations are also known to correlate strongly with halo formation history \citep{ludlow_mass-concentration-redshift_2016} this model could be combined with that described by \cite{benson_halo_2019} to explore correlations between spin, concentration, and environment \citep[as examined by][for example]{johnson_secondary_2019}. We intend to explore these correlations in a future work.

\begin{figure}
 \includegraphics[width=85mm]{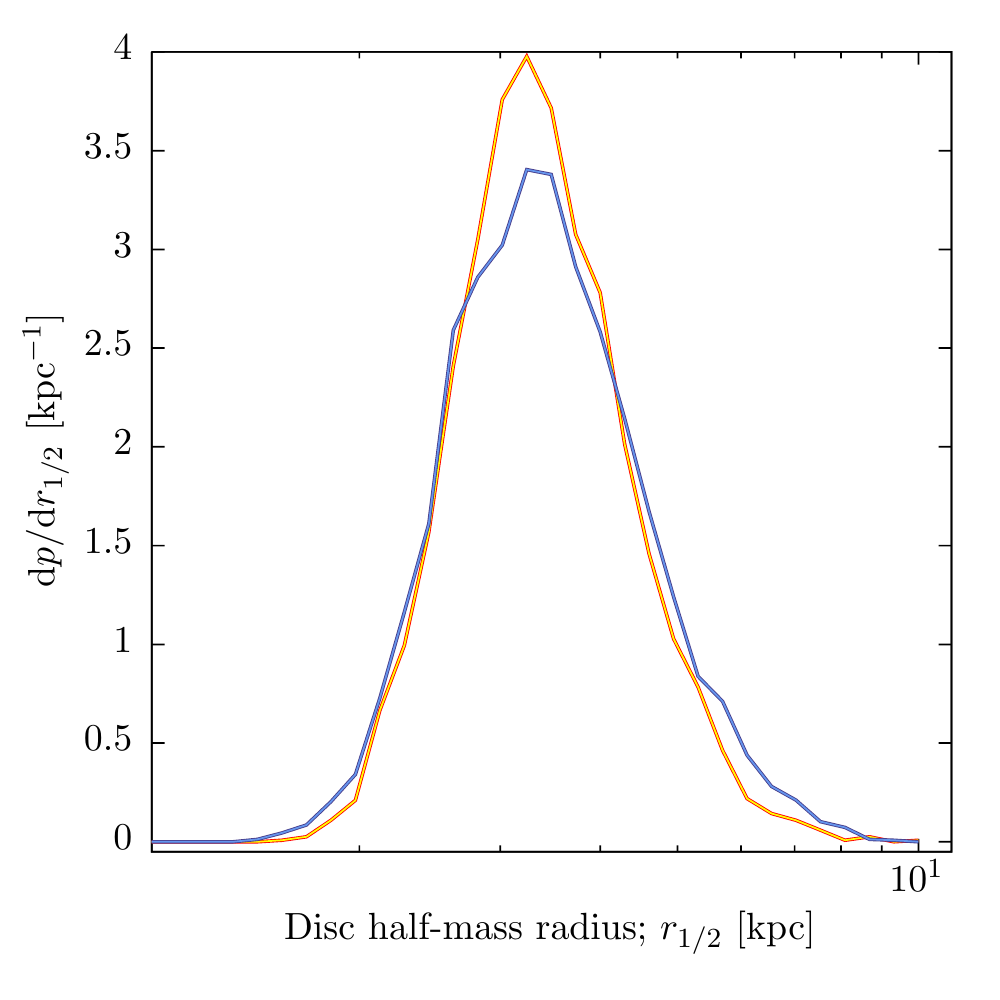}
 \caption{The distribution of half-mass radii of galactic discs in disc-dominated central galaxies occupying $2\times 10^{12}\mathrm{M}_\odot$ halos at $z=0$ as computed using the {\sc Galacticus} semi-analytic model. The yellow line shows results obtained using the Galform model for spins, in which spins are selected at random from a distribution, and updated each time a halo doubles its mass, while the blue line shows results when the model for spins developed in this work is used.}
 \label{fig:sizeDistribution}
\end{figure}

Perhaps most importantly, since this model approximately captures the correlation structure of halo spin (both its magnitude and direction) over time, this opens up the possibility of more accurately tracking angular momenta of halos and galactic discs in semi-analytic models of galaxy formation without the need for very high resolution N-body simulations\footnote{As shown by \protect\cite{benson_constraining_2017} a 10\% precision measurement of $\lambda_\mathrm{P}$ in an N-body halo requires that the halo be resolved with at least 50,000 particles. Since galaxies form over time as their halo grows this means that all progenitor halos in which the galaxy undergoes significant growth must also be resolved with a similarly large number of particles.}, or ad-hoc assumptions about when and how spins change. As a preliminary test of how galaxy sizes might be affected by the use of this model we show in Figure~\ref{fig:sizeDistribution} a comparison of the distribution of half-mass radii of galactic discs in disc-dominated central galaxies in $2\times 10^{12}\mathrm{M}_\odot$ halos at $z=0$ as computed using the {\sc Galacticus} semi-analytic model\footnote{This model for the sizes of galactic disks has its origins in the works of \protect\cite{fall_formation_1980} and \protect\cite{mo_formation_1998} and assumes that the angular momentum content of the galactic disk is directly related to that of the dark matter halo. Unlike those models, in which there is a perfect correlation between the instantaneous angular momentum of the disk and that of the host halo, in {\sc Galacticus} (see also \protect\citealt{cole_hierarchical_2000}) the angular momentum of the disk is accumulated over time as gas cools within each halo, and is further modified by feedback-driven outflows. As such, the disk angular momentum is not perfectly correlated with the instantaneous spin of the halo. Several works \protect\citep{chen_angular_2003,bett_angular_2010,sharma_origin_2012,bryan_impact_2013,lu_formation_2015,liao_segregation_2017,zjupa_angular_2017,jiang_is_2019} have demonstrated significant deviations between the angular momentum conent of dark matter and baryons in a halo. This may suggest that semi-analytic models for the sizes of galactic disks, such as that utilized here, require significant modification. Nevertheless, a reliable model for the spins of dark matter halos is likely to remain an important ingredient for any such improved model.} \citep{benson_galacticus:_2012} using the ``Galform'' model\footnote{In which spins are selected at random, and updated each time a halo doubles its mass, and which has been the standard option in {\sc Galacticus} to date.} for spins \citep{cole_hierarchical_2000}, and using the model developed in this paper\footnote{Note that, when computing the angular momentum of galactic discs (which is the primary determinant of their sizes), {\sc Galacticus} currently ignores only utilizes information about the magnitude of the spin, ignoring the direction of the spin vector---i.e. it effectively assumes that the spin vector always points in the same direction when computing disc angular momentum. This is a simplification made by the model (since, until now, reliable spin vectors have not been available for the majority of halos), which we intend to improve upon as a result of the developments made in this present work. Correctly taking into account the vector nature of spin will likely also change the resulting distribution of disc sizes.}. We find only small changes in the distribution of galaxy sizes when adopting the spin model developed in this work. Specifically the mean logarithm of disc half-mass radius increases from $\langle \log_{10} (r_{1/2}/\mathrm{kpc}) \rangle = 0.529$ to $0.535$, with the root-variance in logarithm of half-mass radius increasing from $\sigma_{\log_{10} (r_{1/2}/\mathrm{kpc}) }=0.106$ to $0.120$. These changes are small, which indicates that previous, simpler models for halo spin should likely not have lead to significantly incorrect results. However, as the model described in this work provides a more detailed description of how the spin of halos evolves, semi-analytic models such as {\sc Galacticus} should be updated to exploit this. For example, {\sc Galacticus} currently does not make use of the vector nature of spin when computing the evolution of disc angular momentum. Accounting for the vector nature of spin will likely also change the resulting distribution of disc sizes. We leave exploration of these consequences of the model developed in this work to a future paper.

\section*{Acknowledgements}

The Millennium Simulation databases used in this paper and the web application providing online access to them were constructed as part of the activities of the German Astrophysical Virtual Observatory (GAVO).

\section*{Data availability}

The data underlying this article are available in Zenodo, at \href{http://doi.org/10.5281/zenodo.3897353}{\tt http://doi.org/10.5281/zenodo.3897353}. N-body simulation data from the Millennium Simulation is publicly available at \href{http://gavo.mpa-garching.mpg.de/MyMillennium/}{\tt http://gavo.mpa-garching.mpg.de/MyMillennium/}.

\bibliographystyle{mn2e}
\bibliography{spins}

\appendix

\section{Tree resolution}\label{sec:resolution}

\begin{figure}
 \includegraphics[width=85mm]{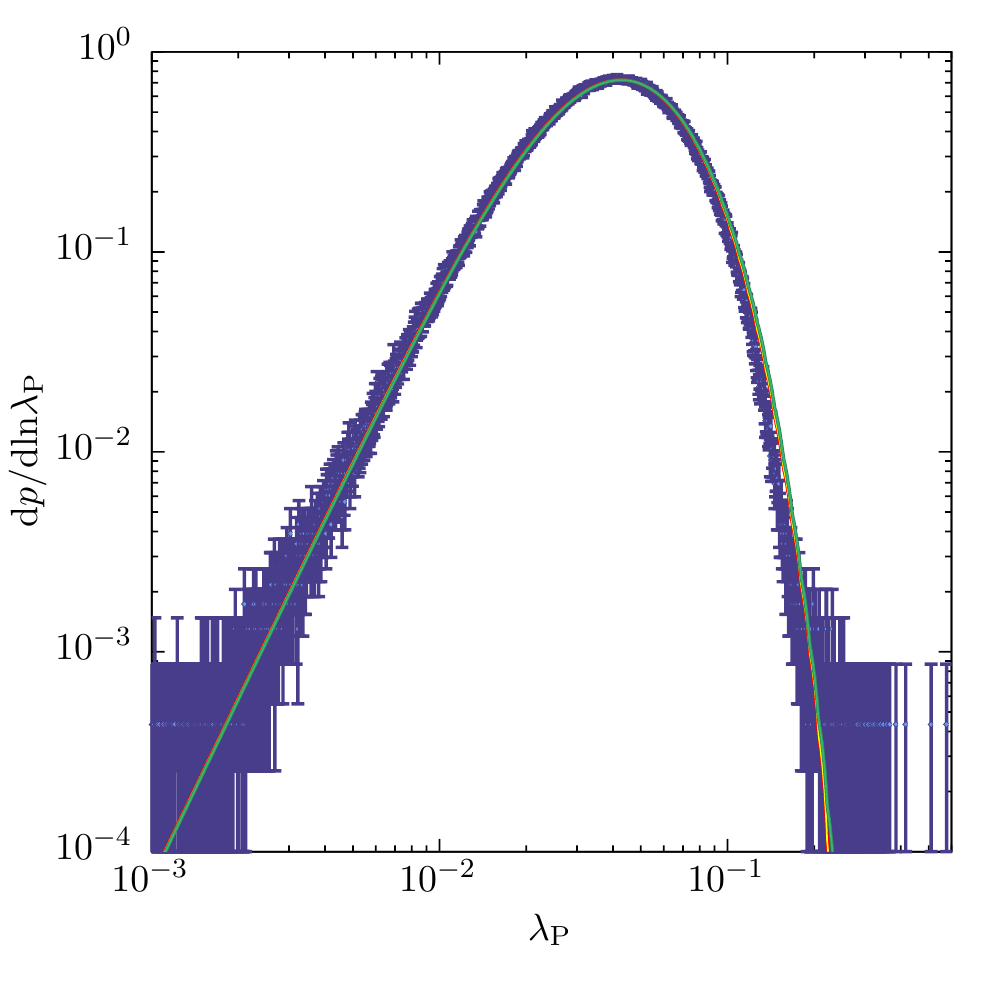}
 \caption{The distribution of spin parameters, $\lambda_\mathrm{P}$, for relaxed halos at $z=0$. The blue points with error bars show the results of \protect\cite{bett_spin_2007}, while the lines show the results from our model for the maximum posterior probability model after convolution with the particle noise distribution predicted by the model of \protect\cite{benson_constraining_2017} for merger tree resolutions of $M_\mathrm{res} = 10^{-3} M_0$ (yellow line), and $M_\mathrm{res} = 10^{-4} M_0$ (green line).}
 \label{fig:spinDistributionHiRes}
\end{figure}

Our merger trees are built with a mass resolution of $M_\mathrm{res} = 10^{-3} M_0$. To test to what extent this finite resolution affects our results we run the maximum posterior probability model with $M_\mathrm{res} = 10^{-4} M_0$. Figure~\ref{fig:spinDistributionHiRes} shows a comparison of this model with the standard resolution of $M_\mathrm{res} = 10^{-3} M_0$ (yellow line), and with the higher resolution of $M_\mathrm{res} = 10^{-4} M_0$ (green line). Using the higher resolution merger trees makes negligible difference to the resulting distribution. Specifically, we find that the distribution of spin parameters for $\lambda_\mathrm{P}$ shifts by less than $0.02$~dex for $\lambda_\mathrm{P} < 0.1$ relative to the $M_\mathrm{res} = 10^{-3} M_0$ case, and shifts by less than $0.05$~dex for $\lambda_\mathrm{P} < 0.2$.

\section{Posterior distribution}\label{sec:posterior}

\begin{figure}
 \newcommand{\triangledir}{plots/modelPosterior}
\renewcommand{\arraystretch}{0}
\setlength{\tabcolsep}{0pt}
\begin{tabular}{l@{}c@{}r@{}l@{}c@{}r@{}}
\multicolumn{3}{c}{\includegraphics[scale=1.0]{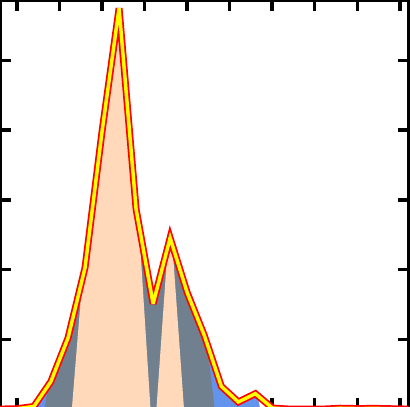}}&\multicolumn{3}{c}{\includegraphics[scale=1.0]{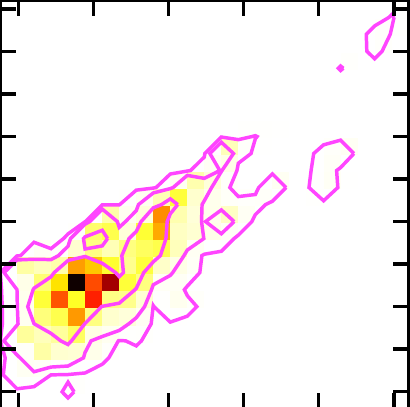}}\\
\multicolumn{1}{p{37.5384615384615pt}}{\raisebox{121pt-\widthof{\normalsize x}-\widthof{\normalsize $-4.0 \times 10^{-2}$}}[0pt][0pt]{\rotatebox{90}{\normalsize $-4.0 \times 10^{-2}$}}}&\multicolumn{1}{p{37.5384615384615pt}}{\hspace{19.6774193548387pt}\raisebox{121pt-\widthof{\normalsize x}-\widthof{\normalsize $ \epsilon$}}[0pt][0pt]{\rotatebox{90}{\normalsize $ \epsilon$}}}&\multicolumn{1}{r}{\raisebox{121pt-\widthof{\normalsize x}-\widthof{\normalsize $9.2 \times 10^{-1}$}}[0pt][0pt]{\rotatebox{90}{\normalsize $9.2 \times 10^{-1}$}\hspace{3pt}}}&\multicolumn{3}{c}{\includegraphics[scale=1.0]{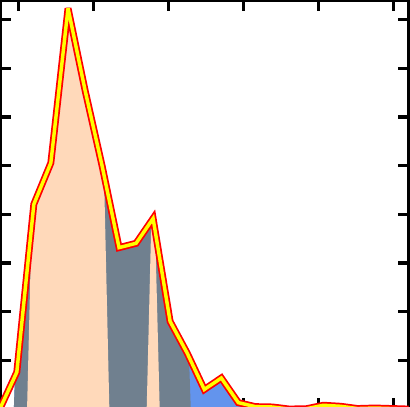}}\\
&&&\multicolumn{1}{p{37.5384615384615pt}}{\raisebox{0pt-\widthof{\normalsize $-2.5 \times 10^{-1}$}}[0pt][0pt]{\rotatebox{90}{\normalsize $-2.5 \times 10^{-1}$}}}&\multicolumn{1}{p{37.5384615384615pt}}{\hspace{19.6774193548387pt}\raisebox{0pt-\widthof{\normalsize $ \alpha$}}[0pt][0pt]{\rotatebox{90}{\normalsize $ \alpha$}}}&\multicolumn{1}{r}{\raisebox{0pt-\widthof{\normalsize $5.2$}}[0pt][0pt]{\rotatebox{90}{\normalsize $5.2$}\hspace{3pt}}}\\
\end{tabular}

 \vspace{15mm}
 \caption{The posterior distribution over the model parameters $\epsilon$ and $\alpha$. (Nuisance parameters are not shown.) The off-diagonal panel shows the posterior distribution over both model parameters, while on-diagonal panels show the posterior distribution over individual model parameters. In the off-diagonal panel, colours show the probability density running from white (low probability density) to dark red (high probability density). Contours are drawn to enclose 99.7\%, 95.4\%, and 68.3\% of the posterior probability when ranked by probability density (i.e. the highest posterior density intervals). In on-diagonal panels the curve indicates the probability density. Shaded regions indicate the 68.3\%, 95.4\%, and 99.7\% highest posterior density intervals.}
 \label{fig:modelPosterior}
\end{figure}

Figure~\ref{fig:modelPosterior} shows the posterior distribution over the model parameters, with nuisance parameters not shown. Both parameters are well-constrained by the N-body data. It is also apparent that their values are strongly correlated in the posterior distribution.

The posterior distribution for $\alpha$ peaks at around $\alpha=0.9$, with a tail extending to $\alpha \approx 3.0$. As the largest spins found for halos are $|\boldsymbol{\lambda}| \approx 0.2$ this means that the largest value expected for $\alpha |\boldsymbol{\lambda}_\mathrm{P}| \approx 0.6 < 1$ such that the probability distribution function described by equation~(\ref{eq:correlatedOrbits}) is always a valid distribution.

\section{Implementation Pseudocode}\label{app:pseudocode}

In this Appendix we provide an implementation of our algorithm for assigning spin parameters to halos in merger trees in pseudo-code, with citations to the relevant equations and source references. We do not describe the merger tree construction algorithm here, but focus solely on the algorithm for assigning spins. This spin algorithm is applicable to any well-formed merger tree. We use the following conventions:
\begin{description}
 \item [$tree$:] the merger tree being processed;
 \item [$halo$:] the current halo being processed;
 \item [$childHalo$:] a child (i.e. progenitor) halo of the current halo;
 \item [$orbit$:] an object describing the orbit of a halo;
 \item [$p_\mathrm{orbit}()$:] a distribution function for orbital parameters;
 \item [$p_\lambda(\lambda)$:] a distribution function for spin parameters;
 \item [$p_\Omega()$:] a distribution function for isotropically-distributed vectors;
 \item [$\boldsymbol{J}_\lambda()$:] the angular momentum as a function of spin parameter and halo properties;
 \item [$\boldsymbol{J}_\mathrm{subres}()$:] a function giving the angular momentum of sub-resolution halos;
 \item [$\leftarrow$:] implies assignment;
 \item [$\cdot firstHalo$:] an operator which returns the first halo for a depth-first walk of a merger tree;
 \item [$\cdot firstChild$:] an operator which returns the first child halo of a halo;
 \item [$\cdot next$:] an operator which returns the next halo in a depth-first walk of a merger tree, or $null$ if no more halos remain;
 \item [$\cdot sibling$:] an operator which returns the next sibling of a halo (i.e. the next halo with the same parent as the operated on halo), or $null$ if no next sibling exists;
 \item [$\cdot sample$:] an operator which samples from a distribution function.
\end{description}

Our implementation is as follows:

\begin{algorithmic}
\State $halo \gets tree\cdot firstHalo$
\While{$halo \ne null$}
 \If{$halo\cdot hasChildren$}
  \State $halo: \boldsymbol{J} \gets 0$
  \State $childHalo \gets halo\cdot firstChild$
  \While{$childHalo \ne null$}
   \State $orbit \gets p_\mathrm{orbit}(halo,childHalo)\cdot sample$ \citep{jiang_orbital_2015}
   \State $halo: \boldsymbol{J} \gets halo: \boldsymbol{J} + orbit\cdot\boldsymbol{J} + childHalo:\boldsymbol{J}$
   \State $childHalo \gets childHalo\cdot sibling$
  \EndWhile
  \State $halo: \boldsymbol{J} \gets halo: \boldsymbol{J} + \boldsymbol{J}_\mathrm{subres}(halo)$ (eqn.~\ref{eq:unresolved})
 \Else
  \State $|\boldsymbol{\lambda}| \gets p_\lambda(\lambda)\cdot sample$ \citep{benson_constraining_2017}
  \State $\hat{\boldsymbol{\lambda}} \gets p_\Omega()\cdot sample$
  \State $halo: \boldsymbol{J} \gets \boldsymbol{J}_\lambda(\boldsymbol{\lambda}, halo)$ (eqn.~\ref{eq:spin})
 \EndIf
 \State $halo \gets halo\cdot next$
\EndWhile
\end{algorithmic}

\end{document}